\documentclass[pra,aps,showpas,floatfix,superscriptaddress,twocolumn,letter]{revtex4}
\usepackage{graphicx}
\usepackage{bm} \usepackage{amssymb}
\usepackage{amsmath} \usepackage{amsfonts}
\usepackage{subfigure}

\newcommand{\ket}[1]{|#1\rangle}
\newcommand{\bra}[1]{\langle#1|}

\newcommand{\dtime}[1]{\frac{\partial#1}{\partial t}}

\begin{document}
\title{Quantum control of dispersion in electromagnetically induced transparency via interacting dressed ground states }
\author{James Owen Weatherall}
\affiliation{Department of Physics and Engineering Physics,
Stevens Institute of Technology, Castle Point on Hudson, Hoboken, NJ 07030, USA}
\affiliation{Department of Mathematical Sciences,
Stevens Institute of Technology, Castle Point on Hudson, Hoboken, NJ 07030, USA}
\affiliation{Department of Logic and Philosophy of Science,
UC Irvine, 3151 Social Science Plaza A, Irvine, CA 92697, USA}
\author{Christopher P. Search}
\affiliation{Department of Physics and Engineering Physics,
Stevens Institute of Technology, Castle Point on Hudson, Hoboken, NJ 07030, USA}

\date{\today}

\begin{abstract}
We offer a general treatment of electromagnetically induced transparency (EIT) in a five level system consisting of four metastable ground states.  Two additional RF/microwave fields coherently couple the two ground states of the standard EIT $\Lambda$ atom to a pair of additional hyperfine states in the ground state manifold, generating two sets of dressed states that interact via the probe and control lasers, which couple to the electronic excited state.  These new hyperfine dressed states manifest themselves in the linear optical susceptibility of the probe as new resonances in addition to the Autler-Townes doublet characteristic of EIT. In particular, we show that the existence of two new narrow resonances, whose width are limited only by ground state decoherence, appear inside the normal EIT transparency window. We show that by controlling the intensity of these RF/microwave fields, one can engineer both the position and width of these narrow resonances and thereby exercise additional control over both the dispersion and group velocity of the probe.
\end{abstract}
\maketitle

\section{Introduction}

Electromagnetically induced transparency (EIT) \cite{harris-first-EIT} is a quantum interference effect in which coherence between atomic states is used to reduce the absorption in a window around an atomic resonance, while simultaneously generating large dispersion and third order nonlinear susceptibilities within the induced transparency window \cite{fleischhauer-review, scully-text}.  The simplest realization of an EIT system consists of a driven $\Lambda$-type atomic system wherein the excited state is coupled to one ground state---the auxiliary state---via a a strong control beam; meanwhile the system is probed by a weak laser beam near resonance with the transition between the excited state and the as-yet uncoupled ground state.

In these simple systems, the characteristic EIT interference effects arise because the control beam induces a splitting of the excited state into a doublet of ``dressed'' states, providing two interfering excitation pathways for the probe beam.  The null in the probe absorption at the normal atomic resonance frequency is a result of destructive quantum interference between these two pathways. The resulting dispersion has very large slope for frequencies within the transparency window around the absorption null, leading to extremely slow probe beam group velocities \cite{lukin+imamoglu, leonhardt-primer}.  Slow light propagation of this sort has been observed in a variety of media, including hot atomic gases \cite{lukin+scully} and atomic Bose-Einstein condensates \cite{hau-first-observation}.

Several groups have expanded on the basic EIT system to further explore the optical properties of coherently prepared atoms.  In general, these modifications to standard EIT have involved coupling additional states to the three levels of standard EIT, via new lasers or RF/microwave fields.  One class of these have led to EIT's sister phenomenon, electromagnetically induced absorption, which occurs when the primary ground state is strongly coupled to an additional, off-manifold state, forming an $N$-configuration \cite{yudin-eia, akulshin-eia}.  Others involve levels on the same electronic manifold as the EIT ground states.  One such modification considers a four level configuration, in which an additional ground state is coupled to the auxiliary state via an RF or optical field.  The resulting doublet constructively interferes with the excited state to produce a new absorption feature in the EIT spectrum, with properties controllable by tuning the Rabi frequency of the additional field \cite{lukin-dark-resonances}.  These so-called ``interacting dark resonances'' have been observed experimentally in several experiments \cite{yu, zhu, wei+manson2}, and they have been suggested for the coherent control of group velocity in the UV regime \cite{mahmoudi-dark-resonances}. Additional theory and experimental work has shown that the feature arising from the interaction between the dark states persists, even in the presence of considerable Doppler broadening \cite{lukin+yelin-doppler, scully-doppler}.

More recently, in an attempt to explain experimental results \cite{wei+manson2}, the absorption spectrum of a similar four level atom was considered, where now the additional state was coupled to the ground state that is probed by the probe laser \cite{wei-autler-townes}.  Other modifications have involved coupling the two grounds states of the $\Lambda$ atom (or other 3-level atomic configurations, such as cascades) to each other, via an RF field \cite{wei+manson1, fleischhauer-full-analytic, korsunsky}.  In these cases, a variety of new phenomena are possible, including efficient frequency conversion for the probe beam.  Alternatively, rather than introduce additional hyperfine levels, recent work has explored the modifications to EIT resulting from coherent tunneling of $\Lambda$ atoms in a double-well Bose-Einstein condensate, forming an effective six-level configuration \cite{weatherall+search-BEC}.  Here, space-like separated atoms can be used to modify the optical response of a sample in one well.

In the current contribution, we consider the more general coherent interactions that occur when both of the $\Lambda$ atom ground states are coherently coupled to additional states.  These interacting dressed states generate new features in the EIT absorption spectrum, whose widths and locations in frequency space can be controlled by independently modulating the Rabi frequencies of the two new coupling fields, providing markedly better control over the optical properties of an atomic sample than possible with previous implementations.  Moreover, we show how, by controlling the intensity of one of the new coupling fields, one can generate group velocities almost 100 times slower than otherwise possible in a given EIT system with a fixed control laser intensity.  Whereas other studies have considered replacing either one or the other ground state with a driven doublet, no general treatment has appeared in which both ground states are driven.  Thus, the phenomenon of the interaction between the doublets---which is what permits the fine tuning of the resonance locations within the transparency window---has yet to be explored.  The ability to tune the relative position of the narrow resonances is what allows for the enhanced control over the dispersion within the transparency window. The current paper includes an analytic statement of the linear susceptibility of the five level atom, from which both the absorption coefficient and dispersion can be simply derived.

The remainder of this paper is organized as follows.  In section \ref{the_model}, we describe our model for the five level atom, dressed by two RF/microwave fields and a strong control laser.  We derive the master equation in a partially dressed basis and show that in the steady state, the coherence term in the master equation relevant to the optical properties of the probe beam permits an analytic solution. In section \ref{properties}, we derive the system's linear susceptibility, $\chi^{(1)}$, from which the optical properties of the sample can be extracted.  Here we also analyze the dispersion and group velocity of the probe laser. Finally in section IV, we comment on present experimental constraints.  An appendix includes the full equations of motion for the relevant terms of the master equation.

\section{The Model}\label{the_model}

We consider a five level atom as depicted in Fig. 1 where two pairs of ground states ($\{\ket{b},\ket{b'}\}$ and $\{\ket{c},\ket{c'}\}$) interact with a single excited state, $\ket{a}$.  Direct transitions between these ground states are assumed to be electric dipole forbidden (ie. $\ket{b}\not\leftrightarrow\ket{c'}$ or $\ket{b}\not\leftrightarrow\ket{c}$, etc.); moreover, to avoid possible degeneracies, we assume that any degeneracy of the states is lifted by an external magnetic field.  In analogy to a standard EIT configuration, a strong laser of field amplitude $\mathcal{E}_{\mu}$ and of fixed frequency $\omega_{\mu}$ propagates near the $\ket{c}\leftrightarrow\ket{a}$ transition.   Here, we study the propagation of a weak probe laser ($\mathcal{E}_p\ll\mathcal{E}_{\mu}$) of variable frequency $\omega_p$ near the $\ket{b}\leftrightarrow\ket{a}$ transition. The couplings between the atomic levels are moderated by their complex Rabi frequencies, which are defined in terms of the lasers driving the transitions.  The $\ket{c}\leftrightarrow\ket{a}$ Rabi frequency, $\Omega_{\mu}$, is given by $\hbar\Omega_{\mu}e^{-i\phi_{\mu}}=\mathcal{E}_{\mu}D_{ac}$ while $\ket{b}\leftrightarrow\ket{a}$ Rabi frequency is similarly given by $\hbar\Omega_{p}e^{-i\phi_{p}}=\mathcal{E}_{p}D_{ab}$.  $D_{ij}=e\bra{i}\mathbf{x}\cdot\epsilon\ket{j}$ is the dipole moment matrix element in the direction of the laser polarization, $\epsilon$, for the $\ket{j}\leftrightarrow\ket{i}$ transition. Additionally, two RF/microwave fields drive magnetic dipole transitions between members of each hyperfine ground state pair: one, of frequency $\omega_{b}$ and Rabi frequency $\Omega_b$ drives the $\ket{b}\leftrightarrow\ket{b'}$ transition; another, of frequency $\omega_{c}$ and Rabi frequency $\Omega_c$ drives the $\ket{c}\leftrightarrow\ket{c'}$. From here on we will refer to the new fields as RF fields for the sake of simplicity. $\Omega_{\mu}$, $\Omega_{b}$, $\Omega_{c}$, and $\Omega_p$ are all taken to be real.

\begin{figure}
\includegraphics[width=.8\columnwidth,angle=270,trim= 0in 1in 2in 1in,clip]{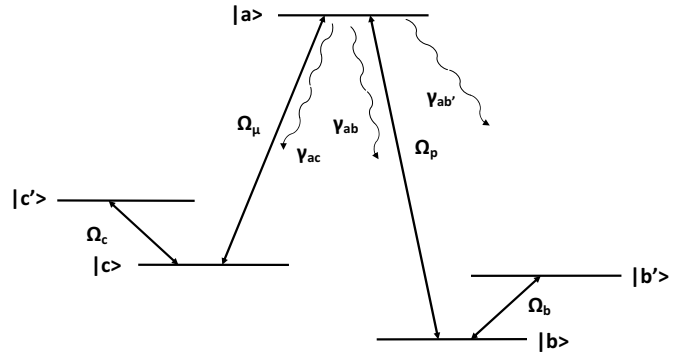}
\caption{\label{Fig_Model}Our 5-level model.  $\ket{b}$, $\ket{b'}$, $\ket{c}$, and $\ket{c'}$ are all in the same ground state manifold; $\ket{a}$ is an excited state.  In analogy to standard EIT, $\ket{a}$ and $\ket{c}$ are coupled by a strong control beam, $\Omega_{\mu}$.  $\ket{b}$ and $\ket{b'}$ and $\ket{c}$ and $\ket{c'}$ are coupled by RF/microwave fields, $\Omega_b$ and $\Omega_c$, respectively.  A weak probe beam, $\Omega_p$, propagates near the $\ket{b}\leftrightarrow\ket{a}$ transition. }
\end{figure}

We take the base vectors for the five atomic levels $\ket{a}$, $\ket{b}$, $\ket{b'}$, $\ket{c}$, and $\ket{c'}$ to form a basis for the relevant subspace of our overall Hilbert space.  Then, in a frame rotating at the frequencies of the fields, defined to avoid explicit time-dependence, we can write the state vector as
\begin{equation}
\ket{\tilde{\Psi}(t)}=\psi_a(t)\ket{a}+\tilde{\psi}_b(t)\ket{b}+\tilde{\psi}_{b'}(t)\ket{b'}+\tilde{\psi}_c(t)\ket{c}+\tilde{\psi}_{c'}(t)\ket{c'},
\end{equation}
where
\begin{subequations}
\begin{align}
\tilde{\psi}_b&=e^{-i\phi_p}e^{-i\nu_p t}\psi_b\\
\tilde{\psi}_{b'}&=e^{-i\phi_p+i\phi_b}e^{-i\nu_p t+i\nu_b t}\psi_{b'}\\
\tilde{\psi}_c&=e^{-i\phi_{\mu}}e^{-i\nu_{\mu} t}\psi_c\\
\tilde{\psi}_{c'}&=e^{-i\phi_{\mu}+i\phi_c}e^{-i\nu_{\mu}t+i\nu_c t}\psi_{c'}.
\end{align}
\end{subequations}
The Hamiltonian in the rotating frame is then given by:
\begin{align}
\tilde{\mathcal{H}}=&\frac{\hbar}{2}\left(\omega_a\ket{a}\bra{a}+(\omega_b+\nu_p)\ket{b}\bra{b}+(\omega_{b'}+\nu_p-\nu_b)\ket{b'}\bra{b'}\right.\notag\\
&+(\omega_c+\nu_{\mu})\ket{c}\bra{c}+(\omega_{c'}+\nu_{\mu}-\nu_c)\ket{c'}\bra{c'}-\Omega_{\mu} \ket{a}\bra{c}\notag\\
&-\Omega_{b}\ket{b'}\bra{b}
-\Omega_{c}\ket{c'}\bra{c}
\left.-\Omega_{p}\ket{a}\bra{b}\right)+\text{h.c.}
\end{align}

To accommodate the possibility of strong fields driving the hyperfine transitions, we will want to keep $\Omega_{b}$ and $\Omega_{c}$ to all orders in our calculation.  To do so, it is convenient to work in a partially dressed basis, in which the subspaces of the effective Hamiltonian corresponding to the two ground state doublets are diagonalized.  We will explicitly diagonalize the $\{\ket{b},\ket{b'}\}$ subspace; the  $\{\ket{c},\ket{c'}\}$ subspace behaves identically mathematically and therefore all results can be inferred from those for the $\{\ket{b},\ket{b'}\}$ subspace.  The $\{\ket{b},\ket{b'}\}$ subspace of the Hamiltonian can be written as a sum of its diagonal and traceless parts:
\begin{equation}
\tilde{\mathcal{H}}_{bb'}=\frac{\hbar}{2}\left((2\omega_b+\Delta_{b}+2\nu_p)\mathbf{I}+
\begin{pmatrix}
-\Delta_{b} &-\Omega_{b}\\
-\Omega_{b} &\Delta_{b}
\end{pmatrix}\right),
\end{equation}
where $\Delta_{b}=\omega_{b'}-\omega_{b}-\nu_b$ is the detuning of the RF field driving the $\ket{b}\leftrightarrow\ket{b'}$ transition.

The matrix diagonalizing $\mathcal{H}_{bb'}$ will be a member of $SO(2)$, so it can be written in terms of a rotation angle in the $\{\ket{b},\ket{b'}\}$ plane, $\theta_b$:
\begin{equation}
D_b=\begin{pmatrix}
\cos\theta_b &\sin\theta_b\\
-\sin\theta_b &\cos\theta_b
\end{pmatrix},
\end{equation}
where
\begin{subequations}
\begin{align}
\cos\theta_b &=\left(\frac{1+\Delta_{b}/\Omega_{b}^{\text{eff}}}{2}\right)^{1/2}\\
\sin\theta_b &=\left(\frac{1-\Delta_{b}/\Omega_{b}^{\text{eff}}}{2}\right)^{1/2}\\
\Omega_{b}^{\text{eff}}&=\sqrt{\Delta_{b}^2+\Omega_{b}^2}.
\end{align}
\end{subequations}
Applying this transformation produces dressed states $\ket{B}=\cos\theta_b\ket{b}+\sin\theta_b\ket{b'}$ and $\ket{B'}=-\sin\theta_b\ket{b}+\cos\theta_b\ket{b'}$ in the diagonalized basis with the energy eigenvalues $\pm \hbar\Omega_b^{eff}/2$.

Combining the transformations for the two subspaces, $D_b$ and $D_c$, into a block diagonal matrix, we find the diagonalization matrix, $D$,
which can be used to determine the partially dress Hamiltonian:
\begin{widetext}
\begin{equation}
D\tilde{\mathcal{H}}D^{\dagger}=\frac{\hbar}{2}
\begin{pmatrix}
2\omega_a &-\Omega_{p}\cos\theta_b &\Omega_p\sin\theta_b &-\Omega_{\mu}\cos\theta_c &\Omega_{\mu}\sin\theta_c\\
-\Omega_{p}\cos\theta_b &2\omega_b+\Delta_{b}+2\nu_p-\Omega_{b}^{\text{eff}} &0 &0 &0\\
\Omega_p\sin\theta_b &0 &2\omega_b+\Delta_{b}+2\nu_p+\Omega_{b}^{\text{eff}} &0 &0\\
-\Omega_{\mu}\cos\theta_c &0 &0 &2\omega_c+\Delta_{c}+2\nu_{\mu}-\Omega_{c}^{\text{eff}} &0\\
\Omega_{\mu}\sin\theta_c &0 &0 &0 &2\omega_c+\Delta_{c}+2\nu_{\mu}+\Omega_{c}^{\text{eff}}
\end{pmatrix}.
\end{equation}
\end{widetext}

At this point we move from a wave function description of the atom to a density matrix, $\tilde{\rho}_{ij}$, that allows us to incorporate decay and decoherence into our model. Since the linear response of the probe is determined by the coherence between the states $\ket{a}$ and $\ket{b}$, $\tilde{\rho}_{ab}=\cos\theta_b\tilde{\rho}_{aB}-\sin\theta_b\tilde{\rho}_{aB'}$, the primary task will be to calculate the steady state solutions of the partially dressed density matrix elements, $\tilde{\rho}_{aB}$ and $\tilde{\rho}_{aB'}$ to linear order in the probe field.

In this paper we consider a closed system in which population from the excited state $|a\rangle$ decays to the included ground states via spontaneous emission and the ground states are stable with respect to decay. Consequently, population is conserved within our 5 level basis and therefore we do not include pumping of the states. Note this restricts our discussion to cold atomic gases where atoms do not quickly pass through the region of interaction with the lasers, which allows us to ignore pumping and decay due to atom transits through the interaction region. For the ground states, we include pure dephasing decoherence between states due to collisions, stray magnetic fields, etc.

The total spontaneous emission rate is given by $\gamma_a$ and we assume that $\ket{a}$ can decay with equal likelihood to $\ket{b}$, $\ket{b'}$, or $\ket{c}$ at the rate $\gamma_a/3$, but because of optical dipole selection rules, it cannot decay to a fourth state. We denote pure dephasing decoherence rates between two states by $\tilde{\gamma}_{ij}$. We make the following assumption that the difference in the dephasing between $\ket{a}$ and both members of the doublet  $\{\ket{b},\ket{b'}\}$ are much smaller than $\gamma_a$, i.e. $|\tilde{\gamma}_{ab}-\tilde{\gamma}_{ab'}|\ll \gamma_a/6$ so that we can set $\gamma_{ab}=\gamma_{ab'}=\gamma_a/6+\tilde{\gamma}_{ab}$ for the total decoherence rate. This approximation is justified by experimental parameters where $\gamma_a \sim 10^{7} s^{-1}$ while collisional decoherences rates can be anywhere from a few tens of $KHz$ down to hundreds of $Hz$ or less for quantum degenerate atomic gases or with buffer gases \cite{pfau,brandt,erhard}. Additionally, we take the decoherence between the dressed state doublets to be of two types, which we allow to vary independently: we have $\tilde{\gamma}_{cb}=\tilde{\gamma}_{cb'}=\gamma_C$ and $\tilde{\gamma}_{c'b}=\tilde{\gamma}_{c'b'}=\gamma_{C'}$.

These two assumptions lead to a considerable simplification of the equations of motion in the partially dressed basis with the relevant contributions being:
\begin{align*}
\dot{\tilde{\rho}}_{aB}&\sim-\gamma_{ab}\tilde{\rho}_{aB}\\
\dot{\tilde{\rho}}_{aB'}&\sim-\gamma_{ab}\tilde{\rho}_{aB'}\\
\dot{\tilde{\rho}}_{CB}&\sim-(\gamma_C\cos^2\theta_c+\gamma_{C'}\sin^2\theta_c)\tilde{\rho}_{CB}&\\&-(\gamma_{C'}-\gamma_{C})\cos\theta_c\sin\theta_c\tilde{\rho}_{C'B}\\
\dot{\tilde{\rho}}_{C'B}&\sim-(\gamma_C\sin^2\theta_c+\gamma_{C'}\cos^2\theta_{c})\tilde{\rho}_{C'B}&\\&-(\gamma_{C'}-\gamma_{C})\cos\theta_c\sin\theta_c\tilde{\rho}_{CB}\\
\dot{\tilde{\rho}}_{CB'}&\sim-(\gamma_C\cos^2\theta_c+\gamma_{C'}\sin^2\theta_c)\tilde{\rho}_{CB'}&\\&-(\gamma_{C'}-\gamma_C)\cos\theta_c\sin\theta_c\tilde{\rho}_{C'B'}\\
\dot{\tilde{\rho}}_{C'B'}&\sim-(\gamma_C\sin^2\theta_c+\gamma_{C'}\cos^2\theta_c)\tilde{\rho}_{C'B'}&\\&-(\gamma_{C'}-\gamma_C)\cos\theta_c\sin\theta_c\tilde{\rho}_{CB'}
\end{align*}
While this second assumption may seem strange, it permits both an analytic solution, and maintains sufficient nuance to allow exploration of the effects of different decoherence pathways on the new resonances as will be shown later.  Note that we do not include $\gamma_{ac}, \gamma_{ac'}$ in this discussion since they do not enter into the solution for $\tilde{\rho}_{ab}$ to first order in the pump field.

We assume that all of the atoms are initially in either $\ket{b}$ or $\ket{b'}$. Since we are solving to linear order in the probe under the assumption $\Omega_p \ll \gamma_{ab}$, many of the terms in the density matrix can be discarded since they only develop population at order $(\Omega_p/\gamma_{ab})^2$ or higher in perturbation theory. For example, $\rho_{aa}\sim\Omega^2_{p}/\gamma^2_{ab}\approx 0$ while $\{\ket{c},\ket{c'}\}$ manifold only develops population at order $\Omega_{p}^2\Omega_{\mu}^2/\gamma_{ab}^4$ which is also negligible. We can then take $\rho_{cc}\approx\rho_{c'c'}\approx\rho_{cc'}\approx\rho_{ac}\approx\rho_{ac'}\approx0$, and likewise their complex conjugates.  Note, however, that the assumption of a weak probe and thus of negligible occupation of $\ket{a}$, $\ket{c}$, and $\ket{c'}$ amounts only to the assumption that $\Omega_p\ll\gamma_{ab}$.  In particular, it places no constraints on the strength of the $\Omega_b$ and $\Omega_c$.  With these assumptions, we find two systems of equations that are decoupled from each other, which we include in the appendix.

Also included in the equations in the appendix are those for the subspace ${\ket{B},\ket{B'}}$, which we note are decoupled from the other equations under the assumption that all terms higher than first order in $\Omega_p$ are negligible, implying that they can be solved separately from the other equations. The terms $\tilde{\rho}_{BB}$, $\tilde{\rho}_{B'B'}$, $\tilde{\rho}_{BB'}$, and $\tilde{\rho}_{B'B}$, which appear in the other equations, act as source terms. In the special case of an on-resonance field driving the $\ket{b}\leftrightarrow\ket{b'}$ transition, we have $\theta_b=\pi/4$ and $\Omega_b^{\text{eff}}=\Omega_b$; thus, Eqs. \ref{bbfull} simplify to:
\begin{align*}
i\dtime{\tilde{\rho}_{BB}}=&-\frac{i\tilde{\gamma}_{bb'}}{2}(\tilde{\rho}_{BB}-\tilde{\rho}_{B'B'})\\
i\dtime{\tilde{\rho}_{B'B'}}=&\frac{i\tilde{\gamma}_{bb'}}{2}(\tilde{\rho}_{BB}-\tilde{\rho}_{B'B'})\\
i\dtime{\tilde{\rho}_{BB'}}=&(-\Omega_b-\frac{i\tilde{\gamma}_{bb'}}{2})\tilde{\rho}_{BB'}+\frac{i\tilde{\gamma}_{bb'}}{2}\tilde{\rho}_{B'B}\\
i\dtime{\tilde{\rho}_{B'B}}=&(\Omega_b-\frac{i\tilde{\gamma}_{bb'}}{2})\tilde{\rho}_{B'B}+\frac{i\tilde{\gamma}_{bb'}}{2}\tilde{\rho}_{BB'}
\end{align*}
Since $\partial\tilde{\rho}_{BB}/\partial t$ and $\partial\tilde{\rho}_{B'B'}/\partial t$ depend only on the difference between $\tilde{\rho}_{BB}$ and $\tilde{\rho}_{B'B'}$, these equations are solved by equal constants---since we have assumed that initially all of the atoms are in the $\{\ket{b},\ket{b'}\}$ manifold, we have $\tilde{\rho}_{BB}=\tilde{\rho}_{B'B'}=1/2$.  As for the coherence terms, we can write their coupled equation of motion as a matrix:
\begin{equation}
i\frac{\partial}{\partial t}\begin{pmatrix}
\tilde{\rho}_{BB'}\\
\tilde{\rho}_{B'B}
\end{pmatrix}
=\left(-\frac{i\tilde{\gamma}_{bb'}}{2}\mathbf{I}+\begin{pmatrix}
-\Omega_b &\frac{i\tilde{\gamma}_{bb'}}{2}\\
\frac{i\tilde{\gamma}_{bb'}}{2} &\Omega_b\end{pmatrix}\right)\begin{pmatrix}
\tilde{\rho}_{BB'}\\
\tilde{\rho}_{B'B}
\end{pmatrix}
\end{equation}
The solution for $\tilde{\rho}_{BB'}$ and $\tilde{\rho}_{B'B}$ can be written as linear combinations of the eigenvectors of the matrix, which have the time dependence $X_1=C_1 e^{(i A-B)t}$ and $X_2=C_2 e^{(-iA-B)t}$ with $A=\sqrt{4(\Omega_b)^2-\gamma^2_{bb'}}/2$ and $B=\gamma_{bb'}/2$.
For $t\gg \gamma_{bb'}$, both of these go to zero, which gives the steady state solutions $\tilde{\rho}_{BB'}=\tilde{\rho}_{B'B}=0$.

To solve for $\tilde{\rho}_{aB}$ and $\tilde{\rho}_{aB'}$, we assume that the $\{\ket{B},\ket{B'}\}$ manifold has been given time to reach the steady state we have described above.  Plugging these into the master equation for $\tilde{\rho}_{aB}$ and $\tilde{\rho}_{aB'}$, and assuming further that the RF coupling $\ket{c}\leftrightarrow\ket{c'}$ transition is also on resonance, $\Delta_c=0$, (giving $\theta_c=\pi/4$), we find the simplified systems:
\begin{subequations}\label{on-resonance-approximation1}
\begin{align}
i\dtime{\tilde{\rho}_{aB}}=&\left(\Delta_p+\frac{\Omega_b}{2}-i\gamma_{ab}\right)\rho_{aB} -\frac{\Omega_p\sqrt{2}}{8} \notag\\ &- \frac{\Omega_{\mu}\sqrt{2}}{4}\left(\tilde{\rho}_{CB}-\tilde{\rho}_{C'B}\right)\\
i\dtime{\tilde{\rho}_{CB}}=&\left(\Delta_{p}-\Delta_{\mu}-\frac{\Omega_c}{2}+\frac{\Omega_b}{2}-\frac{i}{2}(\gamma_{C}+\gamma_{C'})\right)\tilde{\rho}_{CB}\notag\\
&-\frac{\Omega_{\mu}\sqrt{2}}{4}\tilde{\rho}_{aB}
-\frac{i}{2}(\gamma_{C'}-\gamma_C)\tilde{\rho}_{C'B}\\
i\dtime{\tilde{\rho}_{C'B}}=&\left(\Delta_{p}-\Delta_{\mu}+\frac{\Omega_c}{2}+\frac{\Omega_b}{2}-\frac{i}{2}(\gamma_C+\gamma_{C'})\right)\tilde{\rho}_{C'B} \notag\\ &+\frac{\Omega_{\mu}\sqrt{2}}{4}\tilde{\rho}_{aB}-\frac{i}{2}(\gamma_{C'}-\gamma_C)\tilde{\rho}_{CB}
\end{align}
\end{subequations}
and
\begin{subequations}\label{on-resonance-approximation2}
\begin{align}
i\dtime{\tilde{\rho}_{aB'}}=&\left(\Delta_p-\frac{\Omega_b}{2}-i\gamma_{ab}\right)\tilde{\rho}_{aB'}
+\frac{\Omega_p\sqrt{2}}{8}\notag\\  &-\frac{\Omega_{\mu}\sqrt{2}}{4}\left(\tilde{\rho}_{CB'}-\tilde{\rho}_{C'B'}\right)\\
i\dtime{\tilde{\rho}_{CB'}}=&\left(\Delta_{p}-\Delta_{\mu}-\frac{\Omega_c}{2}-\frac{\Omega_b}{2}-\frac{i}{2}(\gamma_C+\gamma_{C'})\right)\rho_{CB'} \notag\\ &-\frac{\Omega_{\mu}\sqrt{2}}{4}\rho_{aB'}-\frac{i}{2}(\gamma_{C'}-\gamma_C)\tilde{\rho}_{C'B'}\\
i\dtime{\tilde{\rho}_{C'B'}}=&\left(\Delta_{p}-\Delta_{\mu}+\frac{\Omega_c}{2}-\frac{\Omega_b}{2}-\frac{i}{2}(\gamma_C+\gamma_{C'})\right)\tilde{\rho}_{C'B'}
\notag\\ &+\frac{\Omega_{\mu}\sqrt{2}}{4}\tilde{\rho}_{aB'} -\frac{i}{2}(\gamma_{C'}-\gamma_C)\tilde{\rho}_{CB'},
\end{align}
\end{subequations}
where $\Delta_p=\omega_a-\omega_b-\nu_p$ and $\Delta_{\mu}=\omega_a-\omega_c-\nu_{\mu}$ are the probe and control beam detunings, respectively.

Eqs. \ref{on-resonance-approximation1} and \ref{on-resonance-approximation2} have identical structures; both can be solved analytically by writing the systems as matrix equations of the form $\partial X/\partial t=-\mathbf{M}\cdot X(t)+A$ and noting that equations of this form have steady state solutions $\lim_{t\leftrightarrow\infty} X(t)=\mathbf{M}^{-1}\cdot A$.  We then extract the steady state solutions to $\tilde{\rho}_{aB}$ and $\tilde{\rho}_{aB'}$ from the resulting vectors.

\section{Optical properties of the 5-level system}\label{properties}

The complex linear susceptibility, expanded about the $\ket{a}\leftrightarrow\ket{b}$ transition, is given by
\begin{equation}
\chi^{(1)}=\frac{2 \sigma(\mathbf{r}) N D_{ab}}{\epsilon_0\mathcal{E}_p}\tilde{\rho}_{ab}.
\end{equation}
$\chi^{(1)}$ determines both the absorption coefficient, $\alpha(\Delta_p)=k_p\Im[\chi^{(1)}(\Delta_p)]$, and the index of refraction, $n(\Delta_p)\approx(1+\Re[\chi^{(1)}(\Delta_p)])^{1/2}$.  Considerations arising from the particular experimental set-up would determine density profile, $\sigma(\mathbf{r})$, and number density of atoms, $N$, which are included to formally account for the particular optical thickness of the sample.
In our analysis we will focus on the reduced susceptibility $\tilde{\chi}^{(1)}=\epsilon_0\hbar\gamma_{ab}\chi^{(1)}/2D_{ab}^2N\sigma(\mathbf{r})=\gamma_{ab}\tilde{\rho}_{ab}/\Omega_p$. We then arrive at the following analytic expression for $\tilde{\chi}^{(1)}$,
\begin{widetext}
\begin{align}
\tilde{\chi}^{(1)}=&\frac{\gamma_{ab}}{2}\left(\frac{(\Delta_{\mu}-\Delta_p+i\gamma_{C'}-\Omega_b/2) (\Delta_{\mu}-\Delta_p+i\gamma_{C}-\Omega_b/2)-\Omega_c^2/4}{(\Delta_p-i\gamma_{ab}+\Omega_b/2) ((\Delta_{\mu}-\Delta_p+i\gamma_{C'}-\Omega_b/2) (\Delta_{\mu}-\Delta_p+i\gamma_{C}-\Omega_b/2)-\Omega_c^2/4)+(\Delta_{\mu}-\Delta_p+i\gamma_{C'}-\Omega_b/2) \Omega_{\mu}^2/4}\right.\notag\\
&\left.+\frac{(\Delta_{\mu}-\Delta_p+i\gamma_{C'}+\Omega_b/2) (\Delta_{\mu}-\Delta_p+i\gamma_{C}+\Omega_b/2)-\Omega_c^2/4}{(\Delta_p-i\gamma_{ab}-\Omega_b/2) ((\Delta_{\mu}-\Delta_p+i\gamma_{C'}+\Omega_b/2) (\Delta_{\mu}-\Delta_p+i\gamma_{C}+\Omega_b/2)-\Omega_c^2/4)+(\Delta_{\mu}-\Delta_p+i\gamma_{C'}+\Omega_b/2) \Omega_{\mu}^2/4}\right)
\end{align}
\end{widetext}
It is worth pointing out that in the limit that $\Omega_b,\Omega_c\rightarrow 0$, we recover the standard expression for the coherence for EIT in a $\Lambda$ system,
\[
\tilde{\chi}^{(1)}\rightarrow\frac{\gamma_{ab}}{2}\frac{\Delta_p-i\tilde{\gamma}_{cb}}{(\Delta_p-i\gamma_{ab})(\Delta_p-i\tilde{\gamma}_{cb})-(\Omega_{\mu}/2)^2}.
\]

Further analysis requires us to estimate the values for the important variables in the problem. The Rabi frequencies of the coupling laser and RF fields are experimentally tunable over a large range. For the spontaneous emission rate, we take $\gamma_a=10^7 s^{-1}$ and will measure the Rabi frequencies and detunings in units of $\gamma_{ab}$. For the ground state dephasing rates the range of values are limited primarily by collision rates and therefore are temperature and density dependent. However for concreteness, we assume that $\tilde{\gamma}_{ab}$, $\gamma_{C}$, and $\gamma_{C'}$ are in the range $10^{3}-10^{4} s^{-1}$.

\begin{figure}
\includegraphics[width=1.0\columnwidth]{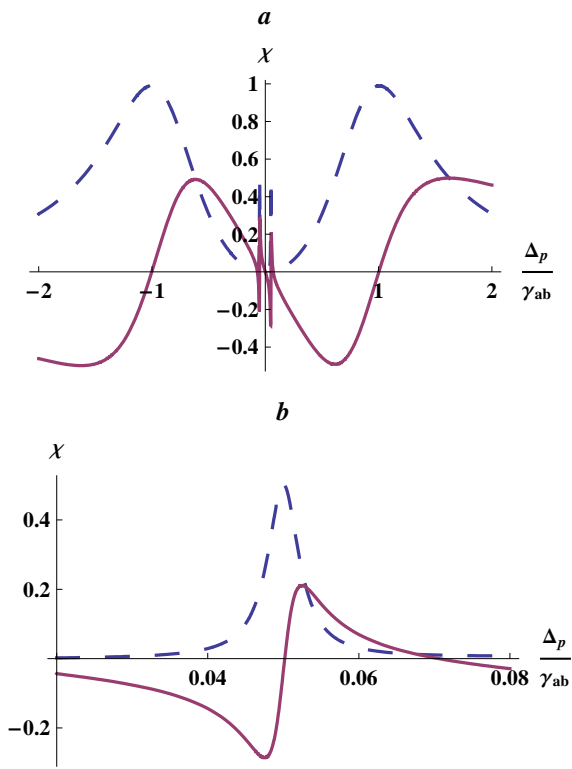}
\caption{\label{new_features}(Color Online) (a) The full spectrum of the susceptibility showing the imaginary part, $\Im[\tilde{\chi}^{(1)}]$, (blue dashed line) and real part $\Re[\tilde{\chi}^{(1)}]$, (red solid line). (b) A close up on one of the new narrow features.  Here, $\Omega_b=\Omega_c=\gamma_{ab}/10$ and $\Omega_{ab}=2\gamma_{ab}$, while  $\gamma_{C}=\gamma_{C'}=0$.}
\end{figure}

\begin{figure}
\includegraphics[width=1.0\columnwidth]{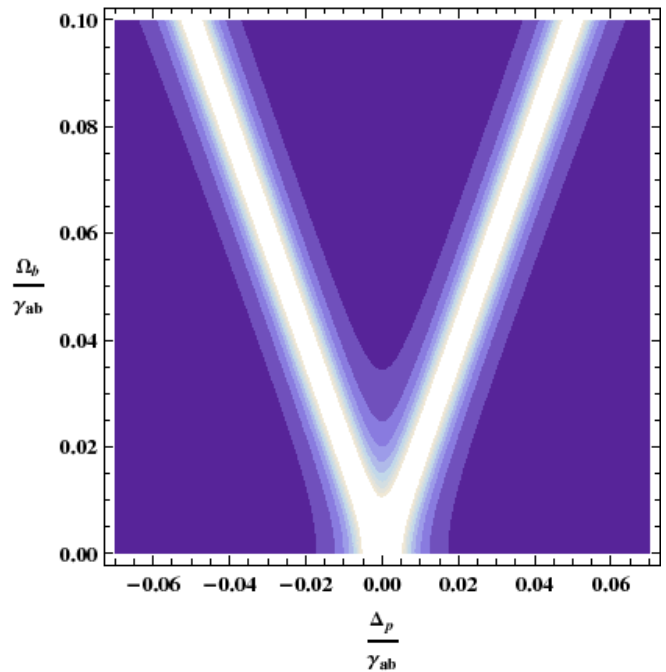}
\caption{\label{fig-varyOb} Imaginary part of the susceptibility, $\Im[\tilde{\chi}^{(1)}]$ plotted as a function of $\Delta_p$ and $\Omega_b$. The separation of the new features varies with $\Omega_b$, the Rabi frequency of the RF field coupling the states $\ket{b}$ and $\ket{b'}$.  Here, we show the dependence for $0\leq\Omega_b\leq\gamma_{ab}/10$, with $\gamma_C=\gamma_{C'}=\gamma_{ab}\times 10^{-3}$, $\Omega_{\mu}=2\gamma_{ab}$, and $\Omega_c=\gamma_{ab}/10$. In the intensity plot, lighter color denotes larger $\Im[\tilde{\chi}^{(1)}]$.}
\end{figure}

Examples of the real and imaginary parts of the susceptibility are shown in Fig. \ref{new_features} for $\Delta_{\mu}=0$ while Fig. \ref{fig-varyOb} display the spectrum's dependence on the Rabi frequency $\Omega_b$. We see that the presence of the additional levels manifest themselves as two narrow resonances located inside of the EIT transparency window. In general for arbitrary $\Delta_b$, the new resonances are symmetrically located about $\Delta_p=0$ at the locations $\Delta_p=\pm \Omega_b^{\text{eff}}/2$. For $\Omega_{\mu},\gamma_{ab}\gg \Omega_b,\Omega_c,\gamma_{C},\gamma_{C'}$, their shape is approximately Lorentzian, given by (for $\Delta_{\mu}=0$):
\begin{equation}
\Im[\tilde{\chi}^{(1)}]\approx\frac{\gamma_{ab}\Omega_c^2}{2\Omega_{\mu}}\left(\frac{\Omega_c^2/\Omega_{\mu}^2+\gamma_{C'}/\gamma_{ab}}{(\Delta_p\mp\Omega_b/2)^2 +(\gamma_{ab}(\Omega_c^2/\Omega_{\mu}^2+\gamma_{C'}/\gamma_{ab}))^2}\right) \label{narrowlorentzian}
\end{equation}
in the vicinity $\Delta_p\approx \pm \Omega_b/2$. In the limit as $\gamma_{C'}\rightarrow 0$, this expression reduces to a Lorentzian with FWHM of $2\gamma_{ab}\Omega_c^2/\Omega_{\mu}^2$, which agrees with what has been obtained by Lukin {\em et al.} \cite{lukin-dark-resonances} and Mahmoudi {\em et al.} \cite{mahmoudi-dark-resonances} for interacting dark resonances in a 4 level system with no ground state dephasing.  In the case of nonzero $\gamma_{C'}$, the widths of the features,
\begin{equation}
\Gamma_n=2\gamma_{ab}(\Omega_c^2/\Omega_{\mu}^2+\gamma_{C'}/\gamma_{ab}), \label{linewidth}
\end{equation}
is the sum of the 'power broadening' term $2\gamma_{ab}\Omega_c^2/\Omega_{\mu}^2$ and the dephasing rate for $|c'\rangle$ while the height is given by
\begin{equation}\label{feature_height}
\Im[\tilde{\chi}^{(1)}(\pm\Omega_b)]=\frac{\Omega_c^2\Omega_{\mu}}{2(\gamma_{ab}\Omega_c^2+\Omega_{\mu}^2\gamma_{C'})}.
\end{equation}

\begin{figure}
\includegraphics[width=1.0\columnwidth]{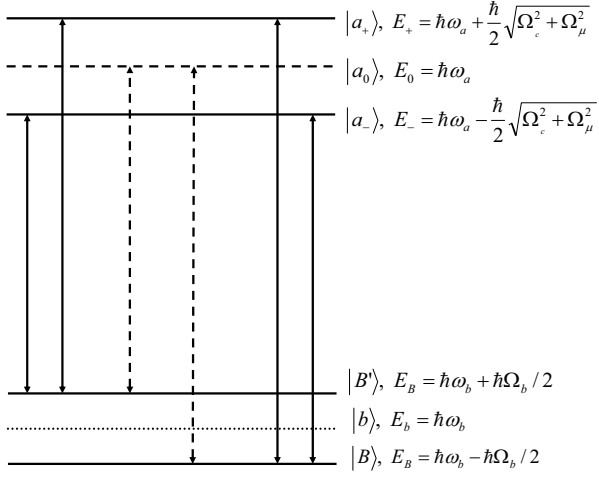}
\caption{\label{dressedstatefig} Energy level diagram that indicates transitions induced by the probe laser between the ground state manifold $\{ |B\rangle, |B'\rangle \}$ and the excited state manifold $\{ |a_+\rangle, |a_-\rangle, |a_0\rangle \}$. Transitions to the dark state $|a_0\rangle$ are indicated by dashed lines. The energy of the bare state $|b\rangle$ is also shown for reference.}
\end{figure}

\begin{figure}
\includegraphics[width=1.0\columnwidth]{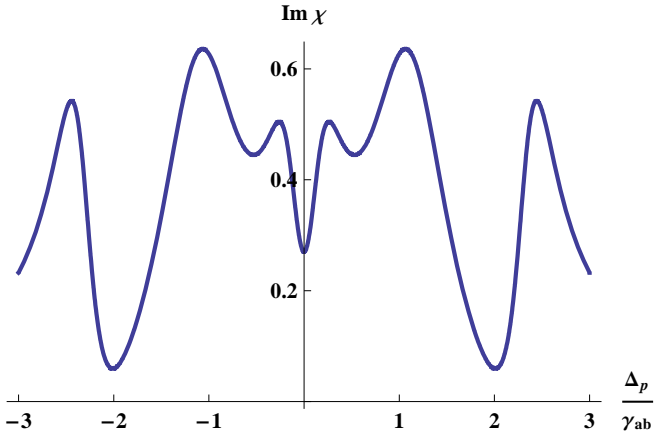}
\caption{\label{six-peaks}Imaginary part of the susceptibility, $\Im[\tilde{\chi}^{(1)}]$, as a function of $\Delta_p$. Here, $\Omega_{\mu}=2\gamma_{ab}$, $\Omega_b=2.2\gamma_{ab}$, $\Omega_c=1.8\gamma_{ab}$ have been chosen so that all six absorption resonances are simultaneously visible.  See text and Fig. \ref{dressedstatefig}.}
\end{figure}

These new resonances have a simple interpretation in terms of the dressed states $|B\rangle$ and $|B'\rangle$, which are coupled via the probe laser to the 3 level ststem  $|c'\rangle \leftrightarrow |c\rangle \leftrightarrow |a\rangle$. The energies of $|B\rangle$ and $|B'\rangle$ are $\hbar\omega_{B,B'}=\hbar\omega_b\pm \Omega_b^\text{eff}/2$ so that in the absence of the control laser, the $|b\rangle \leftrightarrow |a\rangle$ absorption line would be split into an Autler-Townes doublet located at $\omega_{a}- \omega_{B}$ and $\omega_a-\omega_{B'}$. The  $|c'\rangle \leftrightarrow |c\rangle \leftrightarrow |a\rangle$ is isomorphic to a $\Lambda$ atom. Again assuming $\Delta_c=0$ and $\Delta_{\mu}=0$, we then have the following eigenstates for the $\{|a\rangle,|c\rangle,|c'\rangle\}$ subsystem,
\begin{eqnarray}
|a_+\rangle &=&\frac{1}{\sqrt{2}}\left(\sin\theta|a\rangle+|c\rangle+\cos\theta|c'\rangle \right) \\
|a_-\rangle &=&\frac{1}{\sqrt{2}}\left(\sin\theta|a\rangle-|c\rangle+\cos\theta|c'\rangle \right) \\
|a_0\rangle &=& \cos\theta |a\rangle -\sin\theta|c'\rangle
\end{eqnarray}
where $\tan\theta=-\Omega_{\mu}/\Omega_c$. The energies of the states $|a_{\pm}\rangle$ are
$E_{\pm}=\hbar\omega_a \pm \hbar\sqrt{\Omega_{\mu}^2+\Omega_c^2}/2$
while $|a_0\rangle$ has energy $E_0=\hbar\omega_a$. As one can see, $|a_0\rangle$ is the same type of dark state that appears in STIRAP and coherent population trapping. In this case, this induced dark state is a superposition of $|a\rangle$ and $|c'\rangle$ {\em but not} $|c\rangle$ . Since it is decoupled from the control laser, there will not be any destructive quantum interference in the probe absorption for transitions to $|a_0\rangle$.
Transitions from the $\{ |B\rangle, |B'\rangle \}$ manifold to $|a_0\rangle$ will then exhibit absorption {\em resonances} at $\omega_a-\omega_{B,B'}$, which correspond to the new narrow resonances. By contrast, destructive interference created by the control would lead to nulls in the absorption at $\omega_a-\omega_{B,B'}$ when $\Omega_c=0$. Fig. \ref{dressedstatefig} shows a schematic diagram of the energy levels of the dressed ground state manifold $\{|B\rangle, |B'\rangle \}$, which are coupled to all three states of the excited state manifold $\{ |a_+\rangle, |a_-\rangle, |a_0\rangle \}$ via the probe. All in all there are six transitions that should appear as resonances in the absorption spectrum. The transitions to the two `bright' states $|a_\pm\rangle$ correspond to the main absorption peaks located at $\Delta_p\approx \pm\Omega_{\mu}/2$ for $\Omega_{\mu}\gg \Omega_c,\Omega_b$. Notice that each of these resonances actually consist of a pair of resonances separated by a distance $\Omega_b$ but only when $\Omega_b > \gamma_{ab}$ can these pairs be individually resolved as shown in Fig. \ref{six-peaks}.

\begin{figure}
\includegraphics[width=1.0\columnwidth]{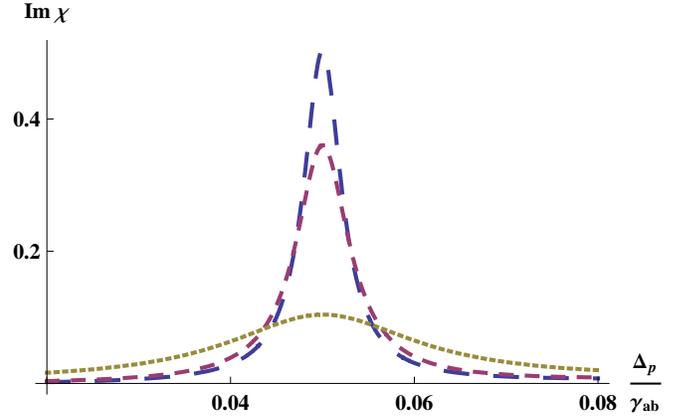}
\caption{\label{varygammacprime} (Color Online) Imaginary part of the susceptibility, $\Im[\tilde{\chi}^{(1)}]$, as a function of $\Delta_p$ showing a close up of the narrow resonances. Here we show the dependence of the new features on the dephasing, $\gamma_{C'}$.  The broad dashes represent the case where $\gamma_{C'}=0$, the short dashed line takes $\gamma_{C'}=\gamma_{ab}\times10^{-3}$, and the dotted line represents $\gamma_{C'}=\gamma_{ab}\times 10^{-2}$.  In all cases, $\Omega_b=\Omega_c=\gamma_{ab}/10$ and $\Omega_{\mu}=2\gamma_{ab}$.}
\end{figure}

The independence of $|a_0\rangle$ from $|c\rangle$ explains why the line widths of the new features only depend on the dephasing between states in the $\{\ket{b},\ket{b'}\}$ manifold and $\ket{a}$ and $\ket{c'}$ but is independent of dephasing relative to state $\ket{c}$ as one can see from Eq. \ref{narrowlorentzian}, which is independent of $\gamma_C$. Fig. \ref{varygammacprime} shows how $\Im[\tilde{\chi}^{(1)}]$ varies with $\gamma_{C'}$.  Moreover, that the height of the features decreases with increasing $\gamma_{C'}$, as in Eq. \ref{feature_height}, reflects the fact that the dephasing decreases the probability of transitions to the dark state $\ket{a_0}$.

These new narrow resonances offer the possibility of additional control of the dispersion, $\partial \Re[\chi^{(1)}]/\partial \omega_p$, and group velocity,
\[
v_g(\omega_p)=\frac{c}{n+(\omega_p/2n)(\partial\Re[\chi^{(1)}]/\partial \omega_p)}
\]
inside of the normal EIT transparency window because of the large normal dispersion in the vicinity of the narrow resonances, whose positions and widths are controlled by $\Omega_b$ and $\Omega_c$, respectively. By controlling the RF Rabi frequencies, the narrow resonances can be selectively positioned inside of the normal transparency window in order to locally control the dispersion over a relatively small frequency range.

To be more specific, we note that in `standard EIT' (note that `standard EIT' refers to the case where $\Omega_b=\Omega_c=0$ but with all other parameters, including the control laser, being the same), the observable effects of EIT including a transparency window and slow light occur when $|\Omega_{\mu}|^2\gg \gamma_{ab}\gamma_{cb}$ but at the same time $\partial \Re[\chi^{(1)}]/\partial \omega_p \propto 2\gamma_{ab}/|\Omega_{\mu}|^2$ for $\Delta_p\approx 0$ \cite{fleischhauer-review}. This implies that although lowering $\Omega_{\mu}$ will decrease the group velocity, the decreasing of the intensity of the control laser to $|\Omega_{\mu}|^2< \gamma_{ab}\gamma_{cb}$ will result in a  complete loss of the transparency window. Let us return now to our 5 level system and focus specifically on the case where $\Omega_{\mu},\gamma_{ab} \gg \Omega_b,\Gamma_n$ so that the narrow resonances are clearly visible in between the Autler-Townes doublet induced by the control laser. In this case the narrow resonances have line widths given by Eq. \ref{linewidth}, which requires that $\Omega_b\gg \Gamma_n$ in order to have a clearly defined window around $\Delta_p=0$. The natural line width of the excited state only contributes through the power broadening term $2(\Omega_c/\Omega_{\mu})^2\gamma_{ab} \ll \gamma_{ab}$ for $\Omega_c\ll \Omega_{\mu}$ and can easily be made much smaller than the ground state dephasing rates. As a result the primary limit on the dispersion and absorption in the middle of the resonances arises from the ground state dephasing alone, requiring $\Omega_b \gg \gamma_{C'}$. This limit is a much weaker condition than for the control laser in standard EIT when $\gamma_{ab}\gg \gamma_{cb}$ and allows for much narrower transparency windows embedded inside of the standard EIT window.

In our system, at $\Delta_p=0$ and assuming $\Omega_b,\Omega_c,\gamma_C,\gamma_{C'}\ll \Omega_{\mu},\gamma_{ab}$, the absorption and dispersion are, to lowest nonvanishing orders of $\Omega_b$, $\Omega_c$, $\gamma_C$, and $\gamma_{C'}$, given by,
\begin{widetext}
\begin{equation}
\Im[\tilde{\chi}^{(1)}(0)]=\gamma_{ab}\frac{4\gamma_{ab}(\Omega_b^2-\Omega_c^2)^2 +(\gamma_C\Omega_b^2+\gamma_{C'}\Omega_c^2)\Omega_{\mu}^2}{\Omega_b^2(\Omega_b^2-\Omega_c^2)^2-2(\Omega_b^4+(\gamma_{ab}\gamma_{C'}+\Omega_b^2)\Omega_c^2)\Omega_{\mu}^2+\Omega_b^2\Omega_{\mu}^4}
\end{equation}
and
\begin{equation}
\partial \Re[\tilde{\chi}^{(1)}]/\partial \Delta_p=\gamma_{ab}\frac{-8\gamma_{ab}(\gamma_C\Omega_b^4+(\gamma_C+3\gamma_{C'})\Omega_b^2\Omega_{c}^2 -\gamma_{C'}\Omega_c^4)\Omega_{\mu}^2-\Omega_b^2(\Omega_b^2+\Omega_c^2)\Omega_{\mu}^4}{\Omega_{\mu}^2(8\gamma_{ab}(\gamma_C\Omega_b^2+\gamma_{C'}\Omega_c^2) +\Omega_b^2(-2\Omega_b^2+\Omega_c^2+\Omega_{\mu}^2))^2}.
\end{equation}
\end{widetext}
From these formulae, in the symmetric case that $\Omega_b=\Omega_c$ and $\gamma_{C}=\gamma_{C'}$, $\Im[\tilde{\chi}^{(1)}(0)]\approx(2\gamma_{C}\gamma_{ab}/\Omega_{\mu}^2)(1+2(\gamma_{ab}\gamma_C+2\Omega_c^2)/\Omega_{\mu}^2)$ while $\partial \Re[\tilde{\chi}^{(1)}]/\partial \Delta_p\approx -(2\gamma_{ab}/\Omega_{\mu}^2)(1-16\gamma_{ab}\gamma_C/\Omega_{\mu}^2)$, which is identical to standard EIT to lowest order. However, slightly off resonance in the vicinity of the narrow resonances where the absorption is still negligible, the affect of these resonances on the dispersion is still very significant.

In Figs. 7 and 8 we consider the effect of the RF Rabi frequencies on the group velocity in the vicinity of the narrow features at $\Delta_p\approx 0$ by plotting the ratio of the group velocity for $\Omega_b,\Omega_c \neq 0$ to the group velocity for standard EIT. Figure 7 shows the group velocity near $\Delta_p=0$ for zero decoherence and small RF Rabi frequencies, $\Omega_b=\Omega_c=0.0001\gamma_{ab}=1KHz$, which leads to an almost 100 fold reduction in the group velocity in a window of about $50Hz$ where the absorption is negligible. Figure 8 shows the reduction of the group velocity and absorption for larger RF Rabi frequencies and finite ground state decoherence. As one can see, even in the presence of decoherence, one can achieve a reduction in the group velocity close to a factor of 10 in a frequency window of $\sim 50KHz$ for the parameters in Fig. 8, while at the same time having relatively small absorption. At the same time, the dispersion is highly nonlinear implying that any pulse propagating near the narrow resonances could experience significant reshaping.

\begin{figure}
\includegraphics[width=1.0\columnwidth]{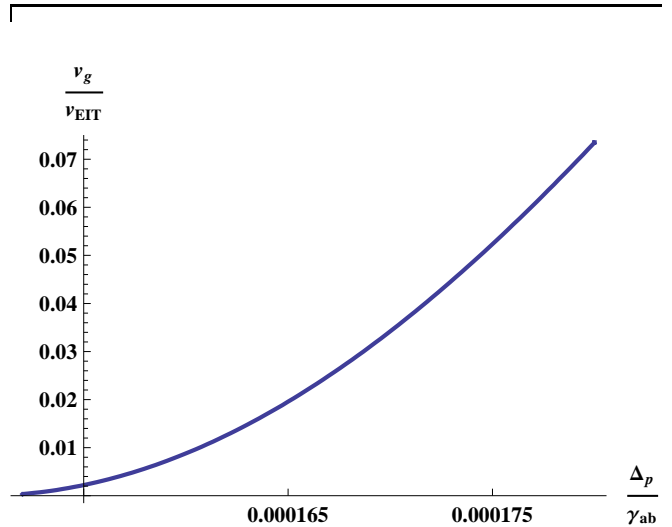}
\caption{\label{fig-groupvelocity} Reduction of group velocity $v_{g}$, in a window near $\Delta_p=0$ for very small RF Rabi frequencies $\Omega_b=\Omega_c=0.0001\gamma_{a}$, $\Omega_\mu=2\gamma_{ab}$, and no decoherence, $\gamma_C=\gamma_{C'}=0$. As in the previous graph, the group velocity is measured relative to the group velocity for $\Omega_b=\Omega_c=0$ but with all other parameters being the same, which we denote as $v_{\text{EIT}}$.}
\end{figure}

\begin{figure}
\includegraphics[width=1.0\columnwidth]{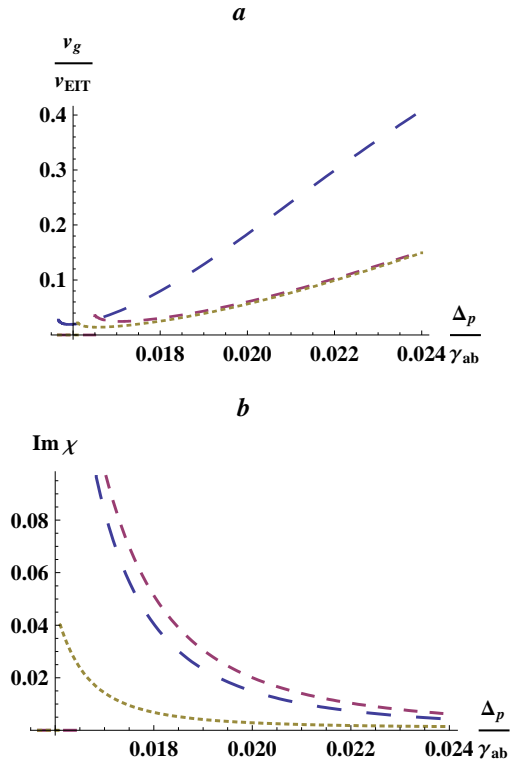}
\caption{\label{fig-groupvelocity} (Color Online) (a) The group velocity, $v_{g}$, in a window near $\Delta_p=0$ for $\Omega_\mu=2\gamma_{ab}$ and  $\Omega_b=\Omega_c=0.01\gamma_{a}$ and $\gamma_C=\gamma_{C'}=0.0001\gamma_{a}$ (Blue long dash line); $\Omega_b=\Omega_c=0.02\gamma_{a}$ and $\gamma_C=\gamma_{C'}=0.0001\gamma_{a}$ (Red short dash line); and $\Omega_b=\Omega_c=0.01\gamma_{a}$ and $\gamma_C=\gamma_{C'}=0$ (Gold dotted line). In each case, the group velocity is measured relative to the group velocity for $\Omega_b=\Omega_c=0$ but with all other parameters being the same, which we denote as $v_{\text{EIT}}$ to mean `standard EIT'. We note that the group velocity can readily be related to the delay time of a light pulse by $\tau_d=L(1/v_{g}-1/c)$ where $L$ is the thickness of the sample. (b) The imaginary part of the susceptibility for the same parameters as in pt. (a).}
\end{figure}

\section{Discussion and conclusion}\label{discussion}
One important aspect that has been ignored in our discussion is that of Doppler broadening. To this end, we note that the current analysis is limited to that of a Doppler free geometry involving co-propagating probe and control lasers \cite{fleischhauer-review}. Doppler shifts will be unimportant for RF/microwave fields. Another alternative to avoid Doppler broadening would be to use laser cooled atoms in a optical trap \cite{pfau}. Additionally, we propose here that the following energy levels of $^{87}$Rb could be used to implement the 5-level geometry (the same transitions but with different electronic number could also be used for $^{23}$Na), $\ket{a}=\ket{5P_{3/2},F=3,m_f=1}$, $\ket{b}=\ket{5S_{1/2},F=2,m_f=2}$, $\ket{b'}=\ket{5S_{1/2},F=2,m_f=1}$, $\ket{c}=\ket{5S_{1/2},F=2,m_f=0}$, and $\ket{c'}=\ket{5S_{1/2},F=1,m_f=0,\pm 1}$ where the probe laser has $\sigma^-$ polarization while the control laser is $\sigma^{+}$ polarized. We note that the five level system could also be implemented using transitions between the $F=1 \rightarrow F'=0$ manifolds, namely: $\ket{a}=\ket{5P_{3/2},F=0,m_f=0}$, $\ket{b}=\ket{5S_{1/2},F=1,m_f=1}$, $\ket{b'}=\ket{5S_{1/2},F=1,m_f=0}$, $\ket{c}=\ket{5S_{1/2},F=1,m_f=-1}$, and $\ket{c'}=\ket{5S_{1/2},F=2,m_f=-2,-1,0}$ again with $\sigma^{-}$ polarized probe and a $\sigma^{+}$ polarized control laser. We note that an additional off-resonant laser could be used to shift the energy level of $|b'\rangle$ relative to $|b\rangle$ and $|c\rangle$ via the AC Stark effect so that an RF field would only resonantly induce spin flips between $|b\rangle$ and $|b'\rangle$ and not between $|b'\rangle$ and $|c\rangle$.

In conclusion we have analyzed the linear susceptibility in a five level system and showed that the emergence of two dark resonances inside of the EIT transparency window offers enhanced control over the dispersion and absorption spectrum due to the controllable widths and separation of the resonances. We have argued that narrower transparency windows and slower group velocities can be achieved using this scheme as compared to standard EIT.

This work is supported in part by the National Science Foundation.

\section{Appendix A}
The full statement of the equations of motion determining $\tilde{\rho}_{aB}$ and $\tilde{\rho}_{aB'}$, approximated by Eqs. \ref{on-resonance-approximation1} and \ref{on-resonance-approximation2} for the case when $\Delta_b=\Delta_c=0$, and so $\theta_b=\theta_c=\pi/4$, are given here in their general form.  We have neglected terms proportional to $\tilde{\rho}_{aa}$, $\tilde{\rho}_{cc}$, $\tilde{\rho}_{c'c'}$, $\tilde{\rho}_{ac}$, $\tilde{\rho}_{ac'}$, and $\tilde{\rho}_{cc'}$, for reasons given in the body of the paper.  We find:
\begin{widetext}
\begin{subequations}
\begin{align}
i\dtime{\tilde{\rho}_{aB}}=&\left(\Delta_p-\frac{\Delta_b}{2}+\frac{\Omega_b^{\text{eff}}}{2}-i\gamma_{ab}\right)\rho_{aB}
-\frac{\Omega_p}{2}\left(\cos\theta_b\rho_{BB}-\sin\theta_b\rho_{B'B}\right)
-\frac{\Omega_{\mu}}{2}\left(\cos\theta_c\rho_{CB}-\sin\theta_c\rho_{C'B}\right)\\
i\dtime{\tilde{\rho}_{CB}}=&\left(\Delta_{p}-\Delta_{\mu}+\frac{\Delta_c}{2}-\frac{\Delta_b}{2}-\frac{\Omega_c^{\text{eff}}}{2}+\frac{\Omega_b^{\text{eff}}}{2}-i(\gamma_C\cos^2\theta_c+\gamma_{C'}\sin^2\theta_c)\right)\rho_{CB} \nonumber \\
-&\frac{\Omega_{\mu}}{2}\cos\theta_c\rho_{aB}-i(\gamma_{C'}-\gamma_{C})\cos\theta_c\sin\theta_c\tilde{\rho}_{C'B}\\
i\dtime{\tilde{\rho}_{C'B}}=&\left(\Delta_{p}-\Delta_{\mu}+\frac{\Delta_c}{2}-\frac{\Delta_b}{2}+\frac{\Omega_c^{\text{eff}}}{2}+\frac{\Omega_b^{\text{eff}}}{2}-i(\gamma_C\sin^2\theta_c+\gamma_{C'}\cos^2\theta_{c})\right)\rho_{C'B} \nonumber \\
+&\frac{\Omega_{\mu}}{2}\sin\theta_c\rho_{aB}-i(\gamma_{C'}-\gamma_{C})\cos\theta_c\sin\theta_c\tilde{\rho}_{CB}
\end{align}
\end{subequations}
and
\begin{subequations}
\begin{align}
i\dtime{\tilde{\rho}_{aB'}}=&\left(\Delta_p-\frac{\Delta_b}{2}-\frac{\Omega_b^{\text{eff}}}{2}-i\gamma_{ab}\right)\rho_{aB'}
+\frac{\Omega_p}{2}\left(\sin\theta_b\rho_{B'B'}-\cos\theta_b\rho_{BB'}\right)
-\frac{\Omega_{\mu}}{2}\left(\cos\theta_c\rho_{CB'}-\sin\theta_c\rho_{C'B'}\right)\\
i\dtime{\tilde{\rho}_{CB'}}=&\left(\Delta_{p}-\Delta_{\mu}+\frac{\Delta_c}{2}-\frac{\Delta_b}{2}-\frac{\Omega_c^{\text{eff}}}{2}-\frac{\Omega_b^{\text{eff}}}{2}-i(\gamma_C\cos^2\theta_c+\gamma_{C'}\sin^2\theta_c)\right)\rho_{CB'} \nonumber \\
-&\frac{\Omega_{\mu}}{2}\cos\theta_c\rho_{aB'}-i(\gamma_{C'}-\gamma_C)\cos\theta_c\sin\theta_c\tilde{\rho}_{C'B'}\\
i\dtime{\tilde{\rho}_{C'B'}}=&\left(\Delta_{p}-\Delta_{\mu}+\frac{\Delta_c}{2}-\frac{\Delta_b}{2}+\frac{\Omega_c^{\text{eff}}}{2}-\frac{\Omega_b^{\text{eff}}}{2}-i(\gamma_C\sin^2\theta_c+\gamma_{C'}\cos^2\theta_c)\right)\rho_{C'B'}
\nonumber \\
+&\frac{\Omega_{\mu}}{2}\sin\theta_c\rho_{aB'}-i(\gamma_{C'}-\gamma_C)\cos\theta_c\sin\theta_c\tilde{\rho}_{CB'}.
\end{align}
\end{subequations}

\begin{subequations} \label{bbfull}
\begin{align}
i\dtime{\tilde{\rho}_{BB}}=&-\frac{i\gamma_{bb'}}{2}(\sin^2\theta_b(\tilde{\rho}_{BB}-\tilde{\rho}_{B'B'})
+\sin2\theta_b\cos2\theta_b(\tilde{\rho}_{BB'}+\tilde{\rho}_{B'B})) \\
i\dtime{\tilde{\rho}_{B'B'}}=&\frac{i\gamma_{bb'}}{2}(\sin^2\theta_b(\tilde{\rho}_{BB}-\tilde{\rho}_{B'B'})
+\sin2\theta_b\cos2\theta_b(\tilde{\rho}_{BB'}+\tilde{\rho}_{B'B})) \\
i\dtime{\tilde{\rho}_{BB'}}=&(-\Omega_b^{\text{eff}}-\frac{i\gamma_{bb'}}{4}(3+\cos4\theta_b))\tilde{\rho}_{BB'}
+\frac{i\gamma_{bb'}}{4}\left((1-\cos4\theta_b)\tilde{\rho}_{B'B} +\sin4\theta_b(\tilde{\rho}_{B'B'}-\tilde{\rho}_{BB})\right) \\
i\dtime{\tilde{\rho}_{B'B}}=&(\Omega_b^{\text{eff}}-\frac{i\gamma_{bb'}}{4}(3+\cos4\theta_b))\tilde{\rho}_{B'B}
+\frac{i\gamma_{bb'}}{4}\left((1-\cos4\theta_b)\tilde{\rho}_{BB'}
+\sin4\theta_b(\tilde{\rho}_{B'B'}-\tilde{\rho}_{BB})\right)
\end{align}
\end{subequations}

\end{widetext}

\bibliography{semiclassical}

\end{document}